\documentclass[     ,final              ]{aipproc}%
\usepackage{amsmath}
\usepackage{amsfonts}
\usepackage{amssymb}
\usepackage{graphicx}%
\layoutstyle{8x11single}
\begin{document}

\title{Running Cosmological Constant and Running Newton Constant in Modified Gravity Theories}


\author{Remo Garattini}{address={Universit\`{a} degli Studi di Bergamo, Facolt\`{a} di Ingegneria,
\\ Viale Marconi 5, 24044 Dalmine (Bergamo) Italy and\\ I.N.F.N. -
sezione di Milano, Milan, Italy.\\ E-mail:remo.garattini@unibg.it}}

\begin{abstract}
We discuss how to extract information about the cosmological constant from the
Wheeler-DeWitt equation, considered as an eigenvalue of a Sturm-Liouville
problem in a de Sitter and Anti-de Sitter background. The equation is
approximated to one loop with the help of a variational approach with Gaussian
trial wave functionals. A canonical decomposition of modes is used to separate
transverse-traceless tensors (graviton) from ghosts and scalar. We show that
no ghosts appear in the final evaluation of the cosmological constant. A zeta
function regularization is used to handle with divergences. A renormalization
procedure is introduced to remove the infinities together with a
renormalization group equation. We apply this procedure on the induced
cosmological constant $\Lambda$ and, as an alternative, on the Newton constant
$G$. A brief discussion on the extension to a $f(R)$ theory is considered.

\end{abstract}
\maketitle

\section{Introduction}

The Friedmann-Robertson-Walker model of the universe, based on the Einstein's
field equations gives an explanation of why the Universe is in an acceleration
phase. However such an expansion must be supported by almost 76\% of what is
known as \textit{Dark Energy}\cite{Obser:2004}. The dark component problem
results from an increasing number of independent cosmological observations,
such as measurements to intermediate and high redshift supernova Ia (SNIa),
measurements of the Cosmic Microwave Background (CMB) anisotropy, and the
current observations of the Large-Scale Structure (LSS) in the universe. The
simplest candidate to explain Dark Energy is based on the equation of state
$P=\omega\rho$ (where $P$ and $\rho$ are the pressure of the fluid and the
energy density, respectively). When $\omega<-1/3$, we are in the Dark energy
regime, while we have a transition to \textit{Phantom Energy} when $\omega
<-1$. In particular, the case of $\omega=-1$ corresponding to a cosmological
constant seems to be a good candidate for the Dark Energy problem. Globally,
this is known as $\Lambda CDM$ model. Nevertheless the $\Lambda CDM$ model
fails in explaining why the observed cosmological constant is so small.
Indeed, there exist 120 order of difference between the estimated cosmological
constant and observation. Basically, the theoretically prediction is based on
the computation of Zero Point Energy (ZPE). One possibility of computing ZPE
in the context of the cosmological constant is given by the Wheeler-DeWitt
equation (WDW)\cite{De Witt:1967}, which is described by%
\begin{equation}
\mathcal{H}\Psi=\left[  \left(  2\kappa\right)  G_{ijkl}\pi^{ij}\pi^{kl}%
-\frac{\sqrt{g}}{2\kappa}\!{}\!\left(  \,\!^{3}R-2\Lambda\right)  \right]
\Psi=0.\label{WDW1}%
\end{equation}
$\kappa=8\pi G$, $G_{ijkl}$ is the super-metric and $^{3}R$ is the scalar
curvature in three dimensions. The main reason to work with a WDW equation
becomes more transparent if we formally re-write the WDW equation
as\cite{Remo:2006}%
\begin{equation}
\frac{1}{V}\frac{\int\mathcal{D}\left[  g_{ij}\right]  \Psi^{\ast}\left[
g_{ij}\right]  \int_{\Sigma}d^{3}x\hat{\Lambda}_{\Sigma}\Psi\left[
g_{ij}\right]  }{\int\mathcal{D}\left[  g_{ij}\right]  \Psi^{\ast}\left[
g_{ij}\right]  \Psi\left[  g_{ij}\right]  }=\frac{1}{V}\frac{\left\langle
\Psi\left\vert \int_{\Sigma}d^{3}x\hat{\Lambda}_{\Sigma}\right\vert
\Psi\right\rangle }{\left\langle \Psi|\Psi\right\rangle }=-\frac{\Lambda
}{\kappa},\label{WDW2}%
\end{equation}
where%
\begin{equation}
V=\int_{\Sigma}d^{3}x\sqrt{g}%
\end{equation}
is the volume of the hypersurface $\Sigma$ and%
\begin{equation}
\hat{\Lambda}_{\Sigma}=\left(  2\kappa\right)  G_{ijkl}\pi^{ij}\pi^{kl}%
-\sqrt{g}^{3}R/\left(  2\kappa\right)  .
\end{equation}
Eq.$\left(  \ref{WDW2}\right)  $ represents the Sturm-Liouville problem
associated with the cosmological constant. In this form the ratio $\Lambda
_{c}/\kappa$ represents the expectation value of $\hat{\Lambda}_{\Sigma}$
without matter fields. The related boundary conditions are dictated by the
choice of the trial wave functionals which, in our case are of the Gaussian
type. Different types of wave functionals correspond to different boundary
conditions. The choice of a Gaussian wave functional is justified by the fact
that we would like to explain the cosmological constant $\left(  \Lambda
_{c}/\kappa\right)  $ as a ZPE effect. To fix ideas, we will work with the
following form of the de Sitter metric (dS)%
\begin{equation}
ds^{2}=-\left(  1-\frac{\Lambda_{dS}}{3}r^{2}\right)  dt^{2}+\frac{dr^{2}%
}{1-\frac{\Lambda_{dS}}{3}r^{2}}+r^{2}\left(  d\theta^{2}+\sin^{2}\theta
d\phi^{2}\right)  \label{dS}%
\end{equation}
and its counterpart the Anti-de Sitter metric (AdS)%
\begin{equation}
ds^{2}=-\left(  1+\frac{\Lambda_{AdS}}{3}r^{2}\right)  dt^{2}+\frac{dr^{2}%
}{1+\frac{\Lambda_{AdS}}{3}r^{2}}+r^{2}\left(  d\theta^{2}+\sin^{2}\theta
d\phi^{2}\right)  ,\label{AdS}%
\end{equation}
which are different expressions of the FRW background. It is interesting to
observe that Eq.$\left(  \ref{WDW2}\right)  $ can be extended to the so-called
modified gravity theories. Basically, one modifies the Einstein-Hilbert action
with the following replacement\cite{f(R):2004}%
\begin{equation}
S=\frac{1}{2\kappa}\int d^{4}x\sqrt{-g}\!R+S^{matter}\quad\rightarrow\quad
S=\frac{1}{2\kappa}\int d^{4}x\sqrt{-g}f\left(  R\right)  +S^{matter}.
\end{equation}
It is clear that other more complicated choices could be done in place of
$f\left(  R\right)  $\cite{CF:2007}. In particular, one could consider
$f\left(  R,R_{\mu\nu}R^{\mu\nu},R_{\alpha\beta\gamma\delta}R^{\alpha
\beta\gamma\delta},\ldots\right)  $ or $f\left(  R,G\right)  $ where $G$ is
the Gauss-Bonnet invariant or any combination of these quantities\footnote{For
a recent riview on $f\left(  R\right)  $, see
Refs.\cite{Faraoni:2008,CF:2007,CDF}, while a recent review on the problem of
$f\left(  G\right)  $ and $f\left(  R,G\right)  $ can be found in
Ref.\cite{NO:2007,ModCosmo:2009}.}. Even if one of the prerogatives of a
$f\left(  R\right)  $ theory is the explanation of the cosmological constant,
we are interested in a more general context, where a combination of $\Lambda$
with a general $f\left(  R\right)  $ theory is considered. As a first step, we
begin to decompose the gravitational perturbation in such a way to obtain the
graviton contribution enclosed in Eq.$\left(  \ref{WDW2}\right)  $.

\section{Extracting the graviton contribution}

We can gain more information if we consider $g_{ij}=\bar{g}_{ij}+h_{ij},$where
$\bar{g}_{ij}$ is the background metric and $h_{ij}$ is a quantum fluctuation
around the background. Thus Eq.$\left(  \ref{WDW2}\right)  $ can be expanded
in terms of $h_{ij}$. Since the kinetic part of $\hat{\Lambda}_{\Sigma}$ is
quadratic in the momenta, we only need to expand the three-scalar curvature
$\int d^{3}x\sqrt{g}{}^{3}R$ up to the quadratic order. However, to proceed
with the computation, we also need an orthogonal decomposition on the tangent
space of 3-metric deformations\cite{Vassilevich:1993,Quad:1969}:%

\begin{equation}
h_{ij}=\frac{1}{3}\left(  \sigma+2\nabla\cdot\xi\right)  g_{ij}+\left(
L\xi\right)  _{ij}+h_{ij}^{\bot}. \label{p21a}%
\end{equation}
The operator $L$ maps $\xi_{i}$ into symmetric tracefree tensors%
\begin{equation}
\left(  L\xi\right)  _{ij}=\nabla_{i}\xi_{j}+\nabla_{j}\xi_{i}-\frac{2}%
{3}g_{ij}\left(  \nabla\cdot\xi\right)  ,
\end{equation}
$h_{ij}^{\bot}$ is the traceless-transverse component of the perturbation
(TT), namely $g^{ij}h_{ij}^{\bot}=0$, $\nabla^{i}h_{ij}^{\bot}=0$ and $h$ is
the trace of $h_{ij}$. It is immediate to recognize that the trace element
$\sigma=h-2\left(  \nabla\cdot\xi\right)  $ is gauge invariant. If we perform
the same decomposition also on the momentum $\pi^{ij}$, up to second order
Eq.$\left(  \ref{WDW2}\right)  $ becomes%
\begin{equation}
\frac{1}{V}\frac{\left\langle \Psi\left\vert \int_{\Sigma}d^{3}x\left[
\hat{\Lambda}_{\Sigma}^{\bot}+\hat{\Lambda}_{\Sigma}^{\xi}+\hat{\Lambda
}_{\Sigma}^{\sigma}\right]  ^{\left(  2\right)  }\right\vert \Psi\right\rangle
}{\left\langle \Psi|\Psi\right\rangle }=-\frac{\Lambda}{\kappa}\Psi\left[
g_{ij}\right]  . \label{lambda0_2}%
\end{equation}
Concerning the measure appearing in Eq.$\left(  \ref{WDW2}\right)  $, we have
to note that the decomposition $\left(  \ref{p21a}\right)  $ induces the
following transformation on the functional measure $\mathcal{D}h_{ij}%
\rightarrow\mathcal{D}h_{ij}^{\bot}\mathcal{D}\xi_{i}\mathcal{D}\sigma J_{1}$,
where the Jacobian related to the gauge vector variable $\xi_{i}$ is%
\begin{equation}
J_{1}=\left[  \det\left(  \bigtriangleup g^{ij}+\frac{1}{3}\nabla^{i}%
\nabla^{j}-R^{ij}\right)  \right]  ^{\frac{1}{2}}.
\end{equation}
This is nothing but the famous Faddev-Popov determinant. It becomes more
transparent if $\xi_{a}$ is further decomposed into a transverse part $\xi
_{a}^{T}$ with $\nabla^{a}\xi_{a}^{T}=0$ and a longitudinal part $\xi
_{a}^{\parallel}$ with $\xi_{a}^{\parallel}=$ $\nabla_{a}\psi$, then $J_{1}$
can be expressed by an upper triangular matrix for certain backgrounds (e.g.
Schwarzschild in three dimensions). It is immediate to recognize that for an
Einstein space in any dimension, cross terms vanish and $J_{1}$ can be
expressed by a block diagonal matrix. Since $\det AB=\det A\det B$, the
functional measure $\mathcal{D}h_{ij}$ factorizes into%
\begin{equation}
\mathcal{D}h_{ij}=\left(  \det\bigtriangleup_{V}^{T}\right)  ^{\frac{1}{2}%
}\left(  \det\left[  \frac{2}{3}\bigtriangleup^{2}+\nabla_{i}R^{ij}\nabla
_{j}\right]  \right)  ^{\frac{1}{2}}\mathcal{D}h_{ij}^{\bot}\mathcal{D}\xi
^{T}\mathcal{D}\psi
\end{equation}
with $\left(  \bigtriangleup_{V}^{ij}\right)  ^{T}=\bigtriangleup
g^{ij}-R^{ij}$ acting on transverse vectors, which is the Faddeev-Popov
determinant. In writing the functional measure $\mathcal{D}h_{ij}$, we have
here ignored the appearance of a multiplicative anomaly\cite{EVZ:1998}. Thus
the inner product can be written as%
\begin{equation}
\int\mathcal{D}h_{ij}^{\bot}\mathcal{D}\xi^{T}\mathcal{D}\sigma\Psi^{\ast
}\left[  h_{ij}^{\bot}\right]  \Psi^{\ast}\left[  \xi^{T}\right]  \Psi^{\ast
}\left[  \sigma\right]  \Psi\left[  h_{ij}^{\bot}\right]  \Psi\left[  \xi
^{T}\right]  \Psi\left[  \sigma\right]  \left(  \det\bigtriangleup_{V}%
^{T}\right)  ^{\frac{1}{2}}\left(  \det\left[  \frac{2}{3}\bigtriangleup
^{2}+\nabla_{i}R^{ij}\nabla_{j}\right]  \right)  ^{\frac{1}{2}}.
\end{equation}
Nevertheless, since there is no interaction between ghost fields and the other
components of the perturbation at this level of approximation, the Jacobian
appearing in the numerator and in the denominator simplify. The reason can be
found in terms of connected and disconnected terms. The disconnected terms
appear in the Faddeev-Popov determinant and these ones are not linked by the
Gaussian integration. This means that disconnected terms in the numerator and
the same ones appearing in the denominator cancel out. Therefore, Eq.$\left(
\ref{lambda0_2}\right)  $ factorizes into three pieces. The piece containing
$\hat{\Lambda}_{\Sigma}^{\bot}$ is the contribution of the
transverse-traceless tensors (TT): essentially is the graviton contribution
representing true physical degrees of freedom. Regarding the vector term
$\hat{\Lambda}_{\Sigma}^{T}$, we observe that under the action of
infinitesimal diffeomorphism generated by a vector field $\epsilon_{i}$, the
components of $\left(  \ref{p21a}\right)  $ transform as
follows\cite{Vassilevich:1993}%
\begin{equation}
\xi_{j}\longrightarrow\xi_{j}+\epsilon_{j},\qquad h\longrightarrow
h+2\nabla\cdot\xi,\qquad h_{ij}^{\bot}\longrightarrow h_{ij}^{\bot}.
\end{equation}
The Killing vectors satisfying the condition $\nabla_{i}\xi_{j}+\nabla_{j}%
\xi_{i}=0,$ do not change $h_{ij}$, and thus should be excluded from the gauge
group. All other diffeomorphisms act on $h_{ij}$ nontrivially. We need to fix
the residual gauge freedom on the vector $\xi_{i}$. The simplest choice is
$\xi_{i}=0$. This new gauge fixing produces the same Faddeev-Popov determinant
connected to the Jacobian $J_{1}$ and therefore will not contribute to the
final value. We are left with%
\begin{equation}
\frac{1}{V}\frac{\left\langle \Psi^{\bot}\left\vert \int_{\Sigma}d^{3}x\left[
\hat{\Lambda}_{\Sigma}^{\bot}\right]  ^{\left(  2\right)  }\right\vert
\Psi^{\bot}\right\rangle }{\left\langle \Psi^{\bot}|\Psi^{\bot}\right\rangle
}+\frac{1}{V}\frac{\left\langle \Psi^{\sigma}\left\vert \int_{\Sigma}%
d^{3}x\left[  \hat{\Lambda}_{\Sigma}^{\sigma}\right]  ^{\left(  2\right)
}\right\vert \Psi^{\sigma}\right\rangle }{\left\langle \Psi^{\sigma}%
|\Psi^{\sigma}\right\rangle }=-\frac{\Lambda}{\kappa}\Psi\left[
g_{ij}\right]  . \label{lambda0_2a}%
\end{equation}
Note that in the expansion of $\int_{\Sigma}d^{3}x\sqrt{g}{}R$ to second
order, a coupling term between the TT component and scalar one remains.
However, the Gaussian integration does not allow such a mixing which has to be
introduced with an appropriate wave functional. Extracting the TT tensor
contribution from Eq.$\left(  \ref{WDW2}\right)  $ approximated to second
order in perturbation of the spatial part of the metric into a background term
$\bar{g}_{ij}$, and a perturbation $h_{ij}$, we get%
\begin{equation}
\hat{\Lambda}_{\Sigma}^{\bot}=\frac{1}{4V}\int_{\Sigma}d^{3}x\sqrt{\bar{g}%
}G^{ijkl}\left[  \left(  2\kappa\right)  K^{-1\bot}\left(  x,x\right)
_{ijkl}+\frac{1}{\left(  2\kappa\right)  }\!{}\left(  \tilde{\bigtriangleup
}_{L\!}\right)  _{j}^{a}K^{\bot}\left(  x,x\right)  _{iakl}\right]  ,
\label{p22}%
\end{equation}
where%
\begin{equation}
\left(  \tilde{\bigtriangleup}_{L\!}\!{}h^{\bot}\right)  _{ij}=\left(
\bigtriangleup_{L\!}\!{}h^{\bot}\right)  _{ij}-4R{}_{i}^{k}\!{}h_{kj}^{\bot
}+\text{ }^{3}R{}\!{}h_{ij}^{\bot} \label{M Lichn}%
\end{equation}
is the modified Lichnerowicz operator and $\bigtriangleup_{L}$is the
Lichnerowicz operator defined by%
\begin{equation}
\left(  \bigtriangleup_{L}h\right)  _{ij}=\bigtriangleup h_{ij}-2R_{ikjl}%
h^{kl}+R_{ik}h_{j}^{k}+R_{jk}h_{i}^{k}\qquad\bigtriangleup=-\nabla^{a}%
\nabla_{a}. \label{DeltaL}%
\end{equation}
$G^{ijkl}$ represents the inverse DeWitt metric and all indices run from one
to three. Note that the term $-4R{}_{i}^{k}\!{}h_{kj}^{\bot}+$ $^{3}R{}%
\!{}h_{ij}^{\bot}$ disappears in four dimensions. The propagator $K^{\bot
}\left(  x,x\right)  _{iakl}$ can be represented as
\begin{equation}
K^{\bot}\left(  \overrightarrow{x},\overrightarrow{y}\right)  _{iakl}%
=\sum_{\tau}\frac{h_{ia}^{\left(  \tau\right)  \bot}\left(  \overrightarrow
{x}\right)  h_{kl}^{\left(  \tau\right)  \bot}\left(  \overrightarrow
{y}\right)  }{2\lambda\left(  \tau\right)  }, \label{proptt}%
\end{equation}
where $h_{ia}^{\left(  \tau\right)  \bot}\left(  \overrightarrow{x}\right)  $
are the eigenfunctions of $\tilde{\bigtriangleup}_{L\!}$. $\tau$ denotes a
complete set of indices and $\lambda\left(  \tau\right)  $ are a set of
variational parameters to be determined by the minimization of Eq.$\left(
\ref{p22}\right)  $. The expectation value of $\hat{\Lambda}_{\Sigma}^{\bot}$
is easily obtained by inserting the form of the propagator into Eq.$\left(
\ref{p22}\right)  $ and minimizing with respect to the variational function
$\lambda\left(  \tau\right)  $. Thus the total one loop energy density for TT
tensors becomes%
\begin{equation}
\frac{\Lambda}{8\pi G}=-\frac{1}{2}\sum_{\tau}\left[  \sqrt{\omega_{1}%
^{2}\left(  \tau\right)  }+\sqrt{\omega_{2}^{2}\left(  \tau\right)  }\right]
. \label{1loop}%
\end{equation}
The above expression makes sense only for $\omega_{i}^{2}\left(  \tau\right)
>0$, where $\omega_{i}$ are the eigenvalues of $\tilde{\bigtriangleup}_{L\!}$.
Concerning the scalar contribution of Eq.$\left(  \ref{lambda0_2a}\right)  $,
in Ref.\cite{Remo1:2004} has been proved that the cosmological constant
contribution is vanishing for a Schwarzschild background. If we follow the
same procedure for dS and AdS metrics, we can show that the only consistent
value is given by $\Lambda_{dS}=\Lambda_{dS}=0$. In the next section, we will
explicitly evaluate Eq.$\left(  \ref{1loop}\right)  $ for a specific background.

\section{One loop energy Regularization and Renormalization for the ordinary
$f\left(  R\right)  =R$ theory}

the dS and AdS metric can be cast into the following form%
\begin{equation}
ds^{2}=-N^{2}\left(  r\left(  x\right)  \right)  dt^{2}+dx^{2}+r^{2}\left(
x\right)  \left(  d\theta^{2}+\sin^{2}\theta d\phi^{2}\right)  ,
\label{metric}%
\end{equation}
where%
\begin{equation}
dx=\pm\frac{dr}{\sqrt{1-\frac{b\left(  r\right)  }{r}}} \label{dx}%
\end{equation}
with%
\begin{equation}
b\left(  r\right)  =\frac{\Lambda_{dS}}{3}r^{3};\qquad b\left(  r\right)
=-\frac{\Lambda_{AdS}}{3}r^{3}.
\end{equation}
$\left(  \tilde{\bigtriangleup}_{L\!}\!{}h^{\bot}\right)  _{ij}$ can be
reduced to%
\begin{equation}
\left[  -\frac{d^{2}}{dx^{2}}+\frac{l\left(  l+1\right)  }{r^{2}}+m_{i}%
^{2}\left(  r\right)  \right]  f_{i}\left(  x\right)  =\omega_{i,l}^{2}%
f_{i}\left(  x\right)  \quad i=1,2\quad, \label{p34}%
\end{equation}
with the help of Regge and Wheeler representation\cite{Regge Wheeler:1957},
where we have used reduced fields of the form $f_{i}\left(  x\right)
=F_{i}\left(  x\right)  /r$ and where we have defined two r-dependent
effective masses $m_{1}^{2}\left(  r\right)  $ and $m_{2}^{2}\left(  r\right)
$%
\begin{equation}
\left\{
\begin{array}
[c]{c}%
m_{1}^{2}\left(  r\right)  =\frac{6}{r^{2}}\left(  1-\frac{b\left(  r\right)
}{r}\right)  +\frac{3}{2r^{2}}b^{\prime}\left(  r\right)  -\frac{3}{2r^{3}%
}b\left(  r\right) \\
\\
m_{2}^{2}\left(  r\right)  =\frac{6}{r^{2}}\left(  1-\frac{b\left(  r\right)
}{r}\right)  +\frac{1}{2r^{2}}b^{\prime}\left(  r\right)  +\frac{3}{2r^{3}%
}b\left(  r\right)
\end{array}
\right.  \quad\left(  r\equiv r\left(  x\right)  \right)  . \label{masses}%
\end{equation}
In order to use the WKB approximation, from Eq.$\left(  \ref{p34}\right)  $ we
can extract two r-dependent radial wave numbers%
\begin{equation}
k_{i}^{2}\left(  r,l,\omega_{i,nl}\right)  =\omega_{i,nl}^{2}-\frac{l\left(
l+1\right)  }{r^{2}}-m_{i}^{2}\left(  r\right)  \quad i=1,2\quad. \label{kTT}%
\end{equation}
To further proceed we use the W.K.B. method used by `t Hooft in the brick wall
problem\cite{tHooft:1985} and we count the number of modes with frequency less
than $\omega_{i}$, $i=1,2$. This is given approximately by%
\begin{equation}
\tilde{g}\left(  \omega_{i}\right)  =\int_{0}^{l_{\max}}\nu_{i}\left(
l,\omega_{i}\right)  \left(  2l+1\right)  dl, \label{p41}%
\end{equation}
where $\nu_{i}\left(  l,\omega_{i}\right)  $, $i=1,2$ is the number of nodes
in the mode with $\left(  l,\omega_{i}\right)  $, such that $\left(  r\equiv
r\left(  x\right)  \right)  $
\begin{equation}
\nu_{i}\left(  l,\omega_{i}\right)  =\frac{1}{\pi}\int_{-\infty}^{+\infty
}dx\sqrt{k_{i}^{2}\left(  r,l,\omega_{i}\right)  }. \label{p42}%
\end{equation}
Here it is understood that the integration with respect to $x$ and $l_{\max}$
is taken over those values which satisfy $k_{i}^{2}\left(  r,l,\omega
_{i}\right)  \geq0,$ $i=1,2$. With the help of Eqs.$\left(  \ref{p41}%
,\ref{p42}\right)  $, Eq.$\left(  \ref{1loop}\right)  $ becomes%
\begin{equation}
\frac{\Lambda}{8\pi G}=-\frac{1}{\pi}\sum_{i=1}^{2}\int_{0}^{+\infty}%
\omega_{i}\frac{d\tilde{g}\left(  \omega_{i}\right)  }{d\omega_{i}}d\omega
_{i}. \label{tot1loop}%
\end{equation}
This is the one loop graviton contribution to the induced cosmological
constant. The explicit evaluation of Eq.$\left(  \ref{tot1loop}\right)  $
gives%
\begin{equation}
\frac{\Lambda}{8\pi G}=\rho_{1}+\rho_{2}=-\frac{1}{4\pi^{2}}\sum_{i=1}^{2}%
\int_{\sqrt{m_{i}^{2}\left(  r\right)  }}^{+\infty}\omega_{i}^{2}\sqrt
{\omega_{i}^{2}-m_{i}^{2}\left(  r\right)  }d\omega_{i},
\end{equation}
where we have included an additional $4\pi$ coming from the angular
integration. The use of the zeta function regularization method to compute the
energy densities $\rho_{1}$ and $\rho_{2}$ leads to%
\begin{equation}
\rho_{i}\left(  \varepsilon\right)  =\frac{m_{i}^{4}\left(  r\right)  }%
{64\pi^{2}}\left[  \frac{1}{\varepsilon}+\ln\left(  \frac{4\mu^{2}}{m_{i}%
^{2}\left(  r\right)  \sqrt{e}}\right)  \right]  \quad i=1,2\quad,
\label{rhoe}%
\end{equation}
where we have introduced the additional mass parameter $\mu$ in order to
restore the correct dimension for the regularized quantities. Such an
arbitrary mass scale emerges unavoidably in any regularization scheme. The
renormalization is performed via the absorption of the divergent part into the
re-definition of a bare classical quantity. Here we have two possible choices:
the induced cosmological constant $\Lambda$ or the gravitational Newton
constant $G$. In addition, we restrict our investigation to the case where%
\begin{equation}
m_{1}^{2}\left(  r\right)  =m_{2}^{2}\left(  r\right)  =m_{0}^{2}\left(
r\right)  , \label{mass}%
\end{equation}
because the dS and AdS backgrounds fall in this case.

\subsection{Running Cosmological Constant}

If we adopt to absorb the divergence using the cosmological constant $\Lambda
$, we can re-define $\Lambda\rightarrow\Lambda_{0}+\Lambda^{div}$, where%
\begin{equation}
\Lambda^{div}=\frac{m_{0}^{4}\left(  r\right)  }{\varepsilon32\pi^{2}}.
\end{equation}
The remaining finite value for the cosmological constant reads%
\begin{equation}
\frac{\Lambda_{0}}{8\pi G}=\left(  \rho_{1}\left(  \mu\right)  +\rho
_{2}\left(  \mu\right)  \right)  =\rho_{eff}^{TT}\left(  \mu,r\right)  ,
\label{lambda0}%
\end{equation}
where $\rho_{i}\left(  \mu\right)  $ has the same form of $\rho_{i}\left(
\varepsilon\right)  $ but without the divergence. The quantity in Eq.$\left(
\ref{lambda0}\right)  $ depends on the arbitrary mass scale $\mu.$ It is
appropriate the use of the renormalization group equation to eliminate such a
dependence. To this aim, we impose that\cite{RGeq:1992}%

\begin{equation}
\frac{1}{8\pi G}\mu\frac{\partial\Lambda_{0}\left(  \mu\right)  }{\partial\mu
}=\mu\frac{d}{d\mu}\rho_{eff}^{TT}\left(  \mu,r\right)  . \label{rg}%
\end{equation}
Solving it we find that the renormalized constant $\Lambda_{0}$ should be
treated as a running one in the sense that it varies provided that the scale
$\mu$ is changing%

\begin{equation}
\frac{\Lambda_{0}\left(  \mu,r\right)  }{8\pi G}=\frac{\Lambda_{0}\left(
\mu_{0},r\right)  }{8\pi G}+\frac{m_{0}^{4}\left(  r\right)  }{16\pi^{2}}%
\ln\frac{\mu}{\mu_{0}}. \label{lambdamu}%
\end{equation}
Substituting Eq.$\left(  \ref{lambdamu}\right)  $ into Eq.$\left(
\ref{lambda0}\right)  $ we find%
\begin{equation}
\frac{\Lambda_{0}\left(  \mu_{0},r\right)  }{8\pi G}=-\frac{1}{32\pi^{2}%
}\left\{  m_{0}^{4}\left(  r\right)  \left[  \ln\left(  \frac{m_{0}^{2}\left(
r\right)  \sqrt{e}}{4\mu_{0}^{2}}\right)  \right]  \right\}  . \label{L0(mu0)}%
\end{equation}
If we go back and look at Eq.$\left(  \ref{WDW2}\right)  $, we note that what
we have actually computed is the opposite of an effective potential (better an
effective energy). Therefore, we expect to find physically acceptable
solutions in proximity of the extrema. We find that Eq.$\left(  \ref{L0(mu0)}%
\right)  $ has an extremum when%
\begin{equation}
\frac{1}{e}=\frac{m_{0}^{2}\left(  \bar{r}\right)  }{4\mu_{0}^{2}}%
\qquad\Longrightarrow\qquad\frac{\bar{\Lambda}_{0}\left(  \mu_{0},\bar
{r}\right)  }{8\pi G}=\frac{m_{0}^{4}\left(  \bar{r}\right)  }{64\pi^{2}%
}=\frac{\mu_{0}^{4}}{4\pi^{2}e^{2}}. \label{LMax}%
\end{equation}
Actually $\bar{\Lambda}_{0}\left(  \mu_{0},\bar{r}\right)  $ is a maximum,
corresponding to a minimum of the effective energy. The effect of the
gravitational fluctuations is to shift the minimum of the effective energy
away from the flat solution leading to an induced cosmological constant.
Plugging Eq.$\left(  \ref{LMax}\right)  $ into Eq.$\left(  \ref{lambdamu}%
\right)  $, we find%
\begin{equation}
\frac{\Lambda_{0}\left(  \mu,r\right)  }{8\pi G}=\frac{\bar{\Lambda}%
_{0}\left(  \mu_{0},\bar{r}\right)  }{8\pi G}+\frac{m_{0}^{4}\left(  r\right)
}{16\pi^{2}}\ln\frac{\mu}{\mu_{0}}=\frac{m_{0}^{4}\left(  \bar{r}\right)
}{64\pi^{2}}\left(  1+4\frac{m_{0}^{4}\left(  r\right)  }{m_{0}^{4}\left(
\bar{r}\right)  }\ln\frac{\mu}{\mu_{0}}\right)
\end{equation}
which can be set to zero when%
\begin{equation}
\frac{\Lambda_{0}\left(  \tilde{\mu},r\right)  }{8\pi G}=0\qquad
\text{\textrm{when\qquad}}\tilde{\mu}=\exp\left(  -\frac{m_{0}^{4}\left(
\bar{r}\right)  }{4m_{0}^{4}\left(  r\right)  }\right)  \mu_{0}.
\label{Lambda0}%
\end{equation}
It is clear that the case is strongly dependent on the background choice. In
this work we fix our attention on dS and AdS metrics written in static way

\subsubsection{dS and AdS background}

In the case of dS and AdS spaces, the effective masses are%
\begin{equation}
m_{1}^{2}\left(  r\right)  =m_{2}^{2}\left(  r\right)  =m_{0}^{2}\left(
r\right)  =\left\{
\begin{array}
[c]{cc}%
\frac{6}{r^{2}}-\Lambda_{dS} & r\in\left(  0,\sqrt{\frac{3}{\Lambda}}\right]
\qquad\text{\textrm{dS Case}}\\
& \\
\frac{6}{r^{2}}+\Lambda_{AdS} & r\in\left(  0,+\infty\right)  \qquad
\text{\textrm{AdS Case}}%
\end{array}
\right.  . \label{masses0}%
\end{equation}
The effective masses have a spurious dependence on $r$, which can be fixed by
the extremum condition described in Eq.$\left(  \ref{LMax}\right)  $. This is
the analogue dependence of the energy momentum tensor on the scale factor $a$
in the Friedmann-Robertson-Walker model. Note that from Eqs.$\left(
\ref{masses0}\right)  $, $m_{0}^{2}\left(  r\right)  $ can never be vanishing,
except for the trivial case of $\Lambda_{dS}=\Lambda_{AdS}=0$. It is
interesting to evaluate the result in proximity of the cosmological throat
$r_{C}=\sqrt{3/\Lambda_{dS}}$ for the dS solution. In this case, Eq.$\left(
\ref{LMax}\right)  $ leads to%
\begin{equation}
\bar{r}^{2}=\frac{6e}{4\mu_{0}^{2}+e\bar{\Lambda}_{dS}}\rightarrow\frac
{4\mu_{0}^{2}}{e}=\bar{\Lambda}_{dS}\qquad\Longrightarrow\qquad\frac
{\bar{\Lambda}_{0}\left(  \mu_{0},\bar{\Lambda}_{dS}\right)  }{8\pi G}%
=\frac{\bar{\Lambda}_{dS}^{2}}{64\pi^{2}}=\frac{\mu_{0}^{4}}{4\pi^{2}e^{2}}%
\end{equation}
and Eq.$\left(  \ref{Lambda0}\right)  $ becomes%
\begin{equation}
\frac{\Lambda_{0}\left(  \tilde{\mu}_{dS},\Lambda_{dS}\right)  }{8\pi
G}=0\qquad\text{\textrm{when\qquad}}\tilde{\mu}_{dS}=\exp\left(  -\frac
{\bar{\Lambda}_{dS}^{2}}{4\Lambda_{dS}^{2}}\right)  \mu_{0}. \label{mdS}%
\end{equation}
For the AdS background, we take the analogue limit of the cosmological throat,
namely $r\rightarrow\infty$, then Eq.$\left(  \ref{LMax}\right)  $ leads to%
\begin{equation}
\bar{r}^{2}=\frac{6e}{4\mu_{0}^{2}-e\bar{\Lambda}_{AdS}}\rightarrow\frac
{4\mu_{0}^{2}}{e}=\bar{\Lambda}_{AdS}\qquad\Longrightarrow\qquad\frac
{\bar{\Lambda}_{0}\left(  \mu_{0},\bar{\Lambda}_{AdS}\right)  }{8\pi G}%
=\frac{\bar{\Lambda}_{AdS}^{2}}{64\pi^{2}}=\frac{\mu_{0}^{4}}{4\pi^{2}e^{2}}%
\end{equation}
and Eq.$\left(  \ref{Lambda0}\right)  $ becomes%
\begin{equation}
\frac{\Lambda_{0}\left(  \tilde{\mu}_{AdS},\Lambda_{AdS}\right)  }{8\pi
G}=0\qquad\text{\textrm{when\qquad}}\tilde{\mu}_{AdS}=\exp\left(  -\frac
{\bar{\Lambda}_{AdS}^{2}}{4\Lambda_{AdS}^{2}}\right)  \mu_{0}. \label{mAdS}%
\end{equation}
Note that at this level of approximation we are unable to distinguish
contributions coming from a dS or AdS background. On the other hand, it is
interesting to observe that if%
\begin{equation}
\Lambda_{dS}\ll\bar{\Lambda}_{dS}\qquad\mathrm{and}\qquad\Lambda_{AdS}\ll
\bar{\Lambda}_{AdS} \label{Lambda}%
\end{equation}
then%
\begin{equation}
\tilde{\mu}_{dS}\ll\mu_{0}\qquad\mathrm{and}\qquad\tilde{\mu}_{AdS}\ll\mu_{0}.
\end{equation}
This means that if we start from $\mu_{0}$ at the Planck scale, we can fine
tune a vanishing cosmological constant for small energy scales. Nevertheless
to obtain this behavior, we need to assume small initial background parameters
in such a way that condition $\left(  \ref{Lambda}\right)  $ is satisfied.

\subsection{Running Newton constant}

If we adopt to absorb the divergence using the Newton constant $G$, we have to
consider the following substitution%
\begin{equation}
\frac{1}{G}\rightarrow\frac{1}{G_{0}\left(  \mu\right)  }+\frac{m_{0}%
^{4}\left(  r\right)  }{\Lambda\varepsilon4\pi}. \label{Ge}%
\end{equation}
Nevertheless, we have to say that this procedure is not immediate for the
Schwarzschild metric and related generalizations. Indeed in this case,
Eq.$\left(  \ref{Ge}\right)  $ becomes%
\begin{equation}
\frac{1}{G}\rightarrow\frac{1}{G_{0}\left(  \mu\right)  }+\left(
\frac{3MG_{0}\left(  \mu\right)  }{r^{3}}\right)  ^{2}\frac{1}{\Lambda
\varepsilon4\pi},
\end{equation}
which means that the divergence is not removed. Therefore, it appears that
this procedure is well defined only for the dS and AdS cases. The remaining
finite value for the cosmological constant reads now%
\begin{equation}
\frac{\Lambda}{8\pi G_{0}\left(  \mu\right)  }=\left(  \rho_{1}\left(
\mu\right)  +\rho_{2}\left(  \mu\right)  \right)  =\rho_{eff}^{TT}\left(
\mu,r\right)  , \label{LG0}%
\end{equation}
where $\rho_{i}\left(  \mu\right)  $ has the same form of $\rho_{i}\left(
\varepsilon\right)  $ but without the divergence. We eliminate the dependence
on the arbitrary mass scale $\mu$, by imposing\cite{RGeq:1992}%

\begin{equation}
\frac{\Lambda}{8\pi}\mu\frac{\partial\left(  G_{0}^{-1}\left(  \mu\right)
\right)  }{\partial\mu}=\mu\frac{d}{d\mu}\rho_{eff}^{TT}\left(  \mu,r\right)
.
\end{equation}
Solving it we find that the renormalized constant $G_{0}$ should be treated as
a running one in the sense that it varies provided that the scale $\mu$ is changing%

\begin{equation}
G_{0}\left(  \mu\right)  =\frac{G_{0}\left(  \mu_{0}\right)  }{1+\frac
{m_{0}^{4}\left(  r\right)  }{32\pi^{2}}G_{0}\left(  \mu_{0}\right)  \ln
\frac{\mu}{\mu_{0}}}. \label{G0}%
\end{equation}
Even in this case, it is interesting to consider the asymptotic part of
$m_{0}^{2}\left(  r\right)  $ for both dS and AdS metrics. It appears from
Eq.$\left(  \ref{G0}\right)  $ that there is a Landau pole at the scale%
\begin{equation}
\left\{
\begin{array}
[c]{c}%
\mu_{0}\exp\left(  -\frac{32\pi^{2}}{\Lambda_{dS}^{2}G_{0}\left(  \mu
_{0}\right)  }\right)  =\mu\qquad\text{\textrm{dS Case}}\\
\\
\mu_{0}\exp\left(  -\frac{32\pi^{2}}{\Lambda_{AdS}^{2}G_{0}\left(  \mu
_{0}\right)  }\right)  =\mu\qquad\text{\textrm{AdS Case}}%
\end{array}
\right.  ,
\end{equation}
invalidating the perturbative calculation\cite{PR:2006,GL:2009}. Substituting
Eq.$\left(  \ref{G0}\right)  $ into Eq.$\left(  \ref{LG0}\right)  $ we find
that the expression of the induced cosmological constant is the same as the
one in Eq.$\left(  \ref{L0(mu0)}\right)  $ with the replacement%
\begin{equation}
\frac{\Lambda_{0}\left(  \mu_{0},r\right)  }{8\pi G}\rightarrow\frac
{\Lambda\left(  r\right)  }{8\pi G_{0}\left(  \mu_{0}\right)  },
\end{equation}
showing that in this case is the Newton constant that is running.
Nevertheless, a fundamental difference in rinormalizing the Newton constant
comes from the fact that we cannot find an appropriate scale where the
cosmological constant can be very small or eventually zero. To this purpose,
we try to generalize this approach including a generic $f\left(  R\right)  $ theory.

\section{Extension to a generic $f\left(  R\right)  $ theory}

It is interesting to note that Eq.$\left(  \ref{WDW2}\right)  $ can be
generalized by replacing the scalar curvature $R$ with a generic function of
$R$. Although a $f\left(  R\right)  $ theory does not need a cosmological
constant, rather it should explain it, we shall consider the following
Lagrangian density describing a generic $f(R)$ theory of gravity
\begin{equation}
\mathcal{L}=\sqrt{-g}\left(  f\left(  R\right)  -2\Lambda\right)
,\qquad{with}\;f^{\prime\prime}\neq0, \label{lag}%
\end{equation}
where $f\left(  R\right)  $ is an arbitrary smooth function of the scalar
curvature and primes denote differentiation with respect to the scalar
curvature. A cosmological term is added also in this case for the sake of
generality, because in any case, Eq.$\left(  \ref{lag}\right)  $ represents
the most general Lagrangian to examine. Obviously $f^{\prime\prime}=0$
corresponds to GR.\cite{Querella:1999}. The semi-classical procedure followed
in this work relies heavily on the formalism outlined in
Refs.\cite{Remo1:2004,CG:2007}. The main effect of this replacement is that at
the scale $\mu_{0}$, we have a shift of the old induced cosmological constant
into%
\begin{equation}
\frac{\Lambda_{0}^{\prime}\left(  \mu_{0},r\right)  }{8\pi G}=\frac{1}%
{\sqrt{h\left(  R\right)  }}\left[  \frac{\Lambda_{0}\left(  \mu_{0},r\right)
}{8\pi G}+\frac{1}{16\pi GV}\int_{\Sigma}d^{3}x\sqrt{g}\frac{Rf^{\prime
}\left(  R\right)  -f\left(  R\right)  }{f^{\prime}\left(  R\right)  }\right]
, \label{Lambdaf(R)}%
\end{equation}
where $V$ is the volume of the system. Note that when $f\left(  R\right)  =R$,
consistently it is $h\left(  R\right)  =1$ with%
\begin{equation}
h\left(  R\right)  =\frac{3f^{\prime}\left(  R\right)  -2}{f^{\prime}\left(
R\right)  } \label{h(R)}%
\end{equation}
We can always choose the form of $f\left(  R\right)  $ in such a way
$\Lambda_{0}\left(  \mu_{0},r\right)  =0$. This implies%
\begin{equation}
\frac{\Lambda_{0}^{\prime}\left(  \mu_{0},r\right)  }{8\pi G}=\frac{1}%
{\sqrt{h\left(  R\right)  }}\frac{1}{16\pi GV}\int_{\Sigma}d^{3}x\sqrt{g}%
\frac{Rf^{\prime}\left(  R\right)  -f\left(  R\right)  }{f^{\prime}\left(
R\right)  }. \label{Lambdaf(R)a}%
\end{equation}
As an example we can examine the following model\footnote{Several models of
$f\left(  R\right)  $ theories are examined in Ref.\cite{AGPT}}%
\begin{equation}
f\left(  R\right)  =AR^{p}\exp\left(  -\alpha R\right)  . \label{f(R)}%
\end{equation}

With this choice, the integrated extra-potential becomes%
\begin{equation}
\frac{1}{V}\int_{\Sigma}d^{3}x\sqrt{g}\frac{Rf^{\prime}\left(  R\right)
-f\left(  R\right)  }{f^{\prime}\left(  R\right)  }=\frac{1}{V}\int_{\Sigma
}d^{3}x\sqrt{g}\frac{R\left(  p-\alpha R-1\right)  }{p-\alpha R}%
\end{equation}
and the function $h\left(  R\right)  $ assumes the form%
\begin{equation}
h\left(  R\right)  =\frac{3A\exp\left(  -\alpha R\right)  R^{p-1}\left(
p-\alpha R\right)  -2}{A\exp\left(  -\alpha R\right)  R^{p-1}\left(  p-\alpha
R\right)  }.
\end{equation}
Note that the scalar curvature is four-dimensional like the argument in
$f\left(  R\right)  $. One can choose, for example the Schwarzschild
background to obtain%
\begin{equation}
\frac{\Lambda_{0}^{\prime}\left(  \mu_{0},r\right)  }{8\pi G}=%
\begin{array}
[c]{c}%
\nearrow\\
\searrow
\end{array}
\left\{
\begin{array}
[c]{cc}%
0 & p\neq0\\
\sqrt{\frac{A\alpha}{3A\alpha+2}}\frac{1}{\alpha16\pi G} & p=0
\end{array}
\right.  .
\end{equation}
while for the dS (AdS) case $R=\pm4\Lambda$ and for the extra-potential one
gets%
\begin{equation}
\frac{1}{V}\int_{\Sigma}d^{3}x\sqrt{g}\frac{R\left(  p-\alpha R-1\right)
}{p-\alpha R}=%
\begin{array}
[c]{c}%
\nearrow\\
\searrow
\end{array}
\left\{
\begin{array}
[c]{cc}%
\frac{4\Lambda\left(  p-\alpha4\Lambda-1\right)  }{p-\alpha4\Lambda} &
\mathrm{dS}\\
\frac{-4\Lambda\left(  p+4\alpha\Lambda-1\right)  }{p+4\alpha\Lambda} &
\mathrm{AdS}%
\end{array}
\right.  . \label{dSAdS}%
\end{equation}
There exists a singularity when $p=\pm\alpha4\Lambda$ for the dS (AdS) case,
while $h\left(  R\right)  $ becomes%
\begin{equation}
h\left(  R\right)  =%
\begin{array}
[c]{c}%
\nearrow\\
\searrow
\end{array}
\left\{
\begin{array}
[c]{cc}%
\frac{3A\exp\left(  -\alpha4\Lambda\right)  \left(  4\Lambda\right)
^{p-1}\left(  p-\alpha4\Lambda\right)  -2}{A\exp\left(  -\alpha4\Lambda
\right)  \left(  4\Lambda\right)  ^{p-1}\left(  p-\alpha4\Lambda\right)  } &
\mathrm{dS}\\
\frac{3A\exp\left(  \alpha4\Lambda\right)  \left(  -4\Lambda\right)
^{p-1}\left(  p+\alpha4\Lambda\right)  -2}{A\exp\left(  \alpha4\Lambda\right)
\left(  -4\Lambda\right)  ^{p-1}\left(  p+\alpha4\Lambda\right)  } &
\mathrm{AdS}%
\end{array}
\right.  .
\end{equation}
From Eq.$\left(  \ref{dSAdS}\right)  $ we can see that%
\begin{equation}
\frac{\Lambda_{0}^{\prime}\left(  \mu_{0},r\right)  }{8\pi G}=0,
\end{equation}
when%
\begin{equation}
\left\{
\begin{array}
[c]{c}%
\Lambda_{dS}=\frac{p-1}{4\alpha}\\
\Lambda_{AdS}=\frac{1-p}{4\alpha}%
\end{array}
\right.  . \label{LdSAdS}%
\end{equation}
Since the modified cosmological constant $\Lambda_{0}^{\prime}$ follows an
evolution equation of the form $\left(  \ref{lambdamu}\right)  $, it appears
that%
\begin{equation}
\frac{\Lambda_{0}^{\prime}\left(  \mu,r\right)  }{8\pi G}=%
\begin{array}
[c]{c}%
\nearrow\\
\searrow
\end{array}
\left\{
\begin{array}
[c]{c}%
\frac{\Lambda_{dS}^{2}}{16\pi^{2}}\ln\frac{\mu}{\mu_{0}}\\
\frac{\Lambda_{AdS}^{2}}{16\pi^{2}}\ln\frac{\mu}{\mu_{0}}%
\end{array}
\right.  \label{Lambdaind}%
\end{equation}
and for the value of $p=1$ for both the dS and AdS background, we get a
vanishing induced cosmological constant also at the scale $\mu$. However, this
result is valid when we assume that the main contribution of the effective
masses is concentrated in proximity of $\Lambda_{dS}$ or $\Lambda_{AdS}$. On
the other hand, when we adopt to renormalize the Newton constant, because of
Eq.$\left(  \ref{Lambdaf(R)a}\right)  $ and Eqs.$\left(  \ref{LdSAdS}\right)
$, the modified cosmological constant cannot be set to zero at any scale,
because there is no a corresponding evolution equation similar to Eq.$\left(
\ref{Lambdaind}\right)  $.

\section{Conclusions}

In this contribution, the effect of a ZPE on the cosmological constant has
been investigated using two specific geometries such as dS and AdS metrics.
The computation has been done by means of a variational procedure with a
Gaussian Wave Functional which should be a good candidate for a ZPE
calculation. We have found that only the graviton is
relevant\cite{GriKos:1989}. Actually, the appearance of a ghost contribution
is connected with perturbations of the shift vectors\cite{Vassilevich:1993}.
In this work we have excluded such perturbations. As usual, in ZPE calculation
we meet the problem of divergences which are regularized with zeta function
techniques. After regularization , we have adopted to remove divergences by
absorbing them into classical quantities: in particular the Newton constant
$G$ and the induced cosmological constant $\Lambda$. This procedure makes
these constants running with the change of the scale $\mu$ appearing in the
regularization scheme. There are two possibilities:

\begin{description}
\item[a)] $\Lambda$ is running. Then we find that it is possible to find some
critical values of the renormalization scale $\mu$ where $\Lambda$ can be set
to zero. However, these points are strongly dependent on the background
choice. The situation changes a little when we replace $R$ with $f\left(
R\right)  $ even if the final result depends on a case to case. In the case
under examination of the model $\left(  \ref{f(R)}\right)  $ for the value of
$p=1$, we find a vanishing cosmological constant at any scale.

\item[b)] $G$ is running. For this case, the induced cosmological constant of
Eq.$\left(  \ref{LG0}\right)  $ cannot be set to zero at any scale. Actually,
the ratio $\Lambda/\left(  8\pi G_{0}\left(  \mu\right)  \right)  $ can be
vanished. Nevertheless, the point where this happens is the Landau point, that
it means that the procedure fails for that value. For a $f\left(  R\right)  $
theory, the problem has not yet examined.
\end{description}

A comment concerning our one loop computation is in order. This approach is
deeply different from the one loop computation of
Refs.\cite{CENOZ:2005,MS:2008}, where the analysis has been done expanding
directly $f\left(  R\right)  $. In our case, the expansion involves only the
three dimensional scalar curvature. Note that with the metric $\left(
\ref{metric}\right)  $ and the effective masses $\left(  \ref{masses}\right)
$, in principle, we can examine every spherically symmetric metric. Note also
the absence of boundary terms in the evaluation of the induced cosmological constant.

\end{document}